\documentstyle[epsfig]{mn}
\input{epsf}
\input rotate
\def\deg{^\circ\!}
\def\bef{\begin{figure}}
\def\eef{\end{figure}}
\def\beq{\begin{equation}}
\def\eeq{\end{equation}}
\def\ber{\begin{eqnarray}}
\def\eer{\end{eqnarray}}
\def\ben{\begin{enumerate}}
\def\een{\end{enumerate}}

\def\beb{\begin{thebibliography}{99}}
\def\eeb{\end{thebibliography}}
\def\bi{\bibitem{}}
\def\etal{{\it et al.}}
\def\eg{{\it e.g. }}
\def\etc{{\it etc}}
\def\ie{{\it i.e. }}

\def\wrt{{\it w.r.t.}}
\def\la{\mathrel{\hbox{\rlap{\hbox{\lower4pt\hbox{$\sim$}}}\hbox{$<$}}}}
\def\ga{\mathrel{\hbox{\rlap{\hbox{\lower4pt\hbox{$\sim$}}}\hbox{$>$}}}}

\def\PASA{{\it Proc. ASA}}
\title[Fluctuation Properties and Polar Emission Mapping of B0834+06] 
{Fluctuation Properties and Polar Emission Mapping
of Pulsar \mbox{B0834+06} at Decameter Wavelengths}
\author[Asgekar \& Deshpande] 
{Ashish Asgekar$^{1,2,3}$
\& Avinash~A. Deshpande$^{1,4}$\\
$^1$Raman Research Institute, Bangalore 560 080 India\\
$^2$Joint Astronomy Programme, Indian Institute of Science,
Bangalore 560 012 India\\
$^3$Department of Physics \& Astronomy, University of Manitoba,  Winnipeg, Manitoba R3T 2N2 Canada\\
$^4$NAIC/Arecibo Observatory, HC3 Box 53995, Arecibo, Puerto Rico 00612 USA\\
ashish@rri.res.in, desh@rri.res.in}
\date{}
\begin{document}
\maketitle

\begin{abstract}
Recent results regarding subpulse-drift in
pulsar \mbox{B0943+10} have led to the identification
of a stable system of sub-beams circulating around the
magnetic axis of the star. Here, we present single-pulse
analysis of pulsar \mbox{B0834+06} at 35~MHz, using observations
from the Gauribidanur Radio Telescope. Certain signatures in the 
fluctuation spectra and correlations allow estimation of
the circulation time and drift direction of the underlying 
emission pattern responsible for the observed modulation. 
We use the `cartographic transform' mapping technique
to study the properties of the polar emission pattern.
These properties are compared with those for the other known case 
of \mbox{B0943+10}, and the implications are discussed.
\end{abstract}

\begin{keywords}
pulsars: general ---
pulsars: individual: B0834+06 ---
radiation mechanism: non-thermal
\end{keywords}

\noindent{\section{Introduction}}

The realization that individual pulses from pulsars display
considerable and often systematic fluctuations 
(Drake \& Craft 1968; Sutton \etal~1970; Backer 1973) followed soon 
after the discovery of pulsars.
Though {\it amplitude modulations} of pulse components were reported
in pulsars with single or multiple components in their average
profiles, more systematic phase modulations
or ``drifting subpulses", occurring in conal single 
(Rankin 1986) pulsars, received a
closer attention from observers and theorists.

Deshpande \& Rankin (1999; hereafter DR99), Deshpande \& Rankin
(2001; hereafter \mbox{Paper-I}), Asgekar \& Deshpande (2001;
hereafter \mbox{Paper-II}) and Rankin, Suleymanova \& Deshpande
(2003; hereafter \mbox{Paper-III}) have reported detailed studies
of drifting subpulses of B0943+10 at frequencies in
the range 35--430 MHz. The studies led to the identification of
a system of sub-beams circulating around the magnetic axis of
\mbox{B0943+10} that was responsible for the observed steady drift
pattern (see DR99).
The 35-MHz \& 103/40 MHz results (\mbox{paper-II} \& \mbox{paper-III}) 
show remarkable similarity with the higher frequency observations.
\mbox{Paper-I} introduced a `cartographic transform'
(and its inverse) which relates the observed pulse sequence to a 
rotating frame around the magnetic axis in order to map the underlying
pattern of emission.  
The implied configuration and the sub-beam circulation
relate well with that envisioned
by Ruderman \& Sutherland (1975; hereafter R\&{S}), wherein
the circulation results from ExB drift.
Whether we observe an amplitude or phase modulation
will, in this picture, depends largely on the viewing 
geometry for a given pulsar.
Although the subpulse modulation in B0943+10 shows
remarkable and rare stability, 
one expects an underlying system of discrete sub-beams 
and their apparent circulation to exist in other pulsars too. 
However, the related details and their possible dependence 
on pulsar parameters are yet to be explored and understood.
Further progress would, therefore, need the identification and 
detailed studies of the emission/subbeam 
patterns to be extended to a sizable sample of pulsars.
Some of the recent investigations on B0834+06 (preliminary results
reported in Asgekar \& Deshpande 2000;
Asgekar 2002), B0809+74 (van Leeuwen \etal~2003), and 
B0826-34 (Gupta \etal~2004) represent attempts in this direction.

Here, we report single-pulse analysis of the pulsar
\mbox{B0834+06} at 34.5~MHz based on a series of 
observations using the Gauribidanur telescope.
This pulsar displays a double profile at frequencies
ranging all the way from 35~MHz to 1~GHz.
It displays strong, alternate-pulse intensity modulation without any
significant drift at meter wavelengths (see, \eg Taylor, Jura
\& Huguenin 1969; Slee \& Mulhall 1970; Sutton \etal~1970, Backer 1973).
It also exhibits frequent, short nulls
(Ritchings 1976). Thus, B0834+06 offers us a useful context
to examine whether the different apparent (single-pulse) modulations, 
particularly the amplitude and phase modulations, share the same 
underlying cause.

\begin{figure*}
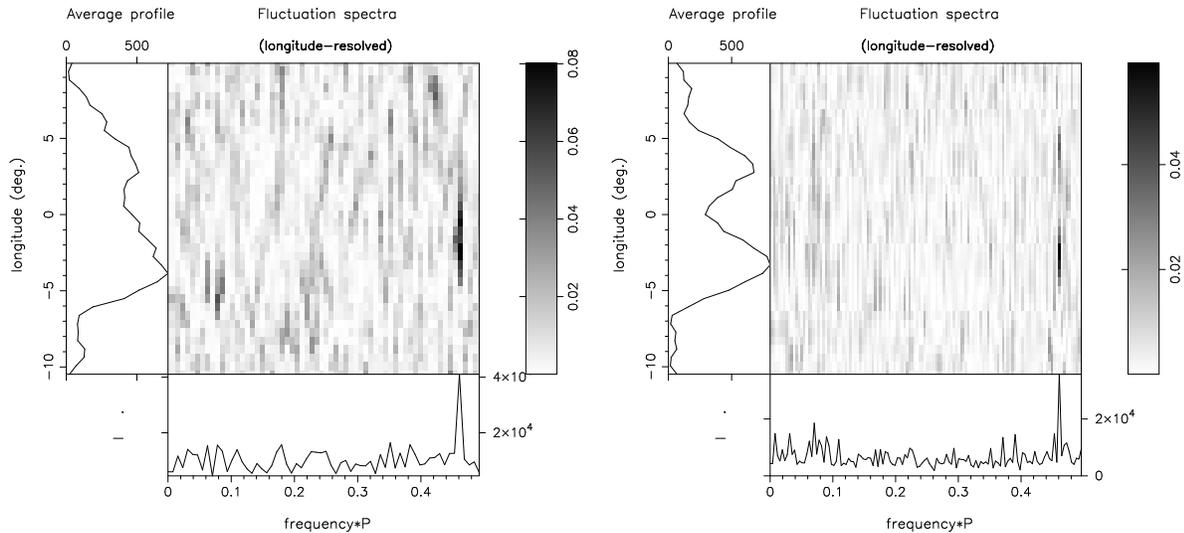

\begin{center}
\begin{tabular}{@{}lr@{}}
{\mbox{\epsfig{file=ME651_FIG_1L.ps,width=7.0cm,angle=-90}}}&
{\mbox{\epsfig{file=ME651_FIG_1R.ps,width=7.0cm,angle=-90}}}\\
\end{tabular}
\end{center}
\caption[LRF-spectra for \mbox{B0834+06}:\mbox{sequence A \& B}]
{Longitude-resolved fluctuation spectra of \mbox{sequences A \& B}.
The strong primary fluctuation feature at 0.46\,c/$P_1$ appears
barely resolved for the B sequence (right panel). The corresponding A-sequence
feature (left panel) is, on the other hand, relatively
broad, indicating a quasi-periodic modulation.}
\label{fig:0834B215_lr}
\end{figure*}

Our observations and analysis procedures are outlined in the
next section. The subsequent sections describe the results
of fluctuation-spectral analysis and our basis for
estimating the true frequencies of the modulation features in
the spectra as well as the period and direction of the
subbeam-system circulation.
In the end, we present the derived polar
emission maps and discuss
our results along with their implications.\\

\noindent{\section{Observations and Analysis}}

Observations of 
B0834+06 were a part of a recent
programme of pulsar observations (Asgekar \& Deshpande 1999)
initiated at the Gauribidanur Radio Observatory (Deshpande,
Shevgaonkar \& Sastry 1989) with a new data acquisition
system for direct voltage-sampling (Deshpande et al. 2004).
The pulsar was observed for
a typical duration of $\ga{1000}$~s in each session. Several
such observations were made during early 1999 and 2002. We
discuss three data sets on B0834+06 below, denoted as `A', `B' and `C',
which were observed on February~5 \&~6, 1999 and Feb. 5, 2002 respectively.
Most of the analysis on sequence~B consisted of 256~pulses, whereas
that of sequence~A \& C consisted of 128~pulses, unless mentioned
otherwise.

In \mbox{paper-II} we detailed our data acquisition system
and reduction procedure used, and our techniques of spectral
analysis are virtually identical to those used in \mbox{Paper-I}.
Readers may refer to these papers for fuller descriptions of the
above aspects.

\noindent{\section{Fluctuation Spectra}

Longitude-resolved fluctuation (LRF) spectra (see
Backer 1973) for the sequences A \& B on \mbox{B0834+06} are shown in 
Figure~\ref{fig:0834B215_lr}. Both spectra show a strong
feature at $\sim0.47$\,c/$P_1$ (hereafter referred as
\mbox{feature-1}), where $P_1$ denotes the rotation period of the pulsar. 
The feature appears barely resolved in the B-sequence spectrum,
indicating a $Q$-value of~$\sim100$, found to be the highest
among Qs seen in our pulse sequences. Sequence A, 
where the periodicity appears to be less stable, is  more
typical. 
This feature represents the alternate-pulse intensity modulation observed
in this pulsar at meter wavelengths.
The frequent, short nulls (typically 1 or 2~null
pulses over $\sim10-20$~pulses) may affect the
$Q$-value of the modulation, as was noted by
Lyne \& Ashworth (1983) for pulsar \mbox{B0809+74}.
However, due to inadequate sensitivity, we can not even
assess, let alone quantify, the possible presence and 
effect of ``nulls" in our data.
The centroid frequency of \mbox{feature-1} is slightly different
for the two data sets, but such small variations ($\sim1\%$)
appear similar to those observed in \mbox{B0943+10}
(\mbox{Paper-I}, \mbox{Paper-II} \& \mbox{Paper-III}).

\begin{figure}
\centerline{\epsfig{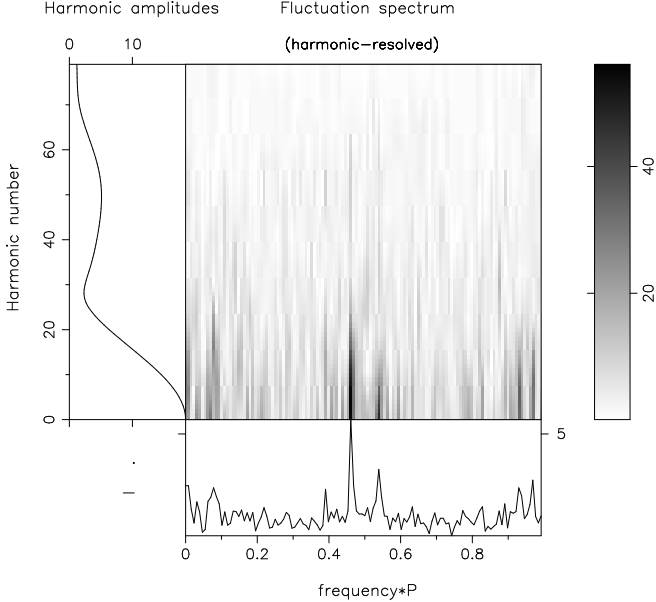}}
\caption[Hrf-spectrum for \mbox{sequence B}]
{HRF-spectrum for \mbox{sequence B} data. A pair of
features placed symmetrically around $0.5\,c/P_1$ are seen. Also
seen are two sidebands on the primary fluctuation feature, which
indicate the presence of a slower modulation in the data modulating
the primary fluctuation (refer to text for details).}
\label{fig:0834D216_hr}
\end{figure}

To examine the issue of possible aliasing of the
spectral features here,
we compute harmonic-resolved fluctuation
spectra (hereafter, ``HRF-spectra", see \mbox{paper-I}:
Figure~4) for the two pulse sequences A and~B. An
HRF-spectrum of a pulse sequence is computed by Fourier
transforming the continuous time series, which is
reconstructed from the (``gated") pulse sequence
by using the available samples on the pulse and filling the
unsampled region with zeros. Such 
spectra can be computed for suitable blocks of 128~or
256~pulses and then the power spectra are averaged. We present, in
Figure~\ref{fig:0834D216_hr}, the HRF-spectrum of the
B~sequence.
The spectrum displays a pair of strong peaks
at about 0.47 and 0.53\,c/$P_1$. 
A similar spectrum for the A~sequence (not shown)
contains just these two features, but of about equal intensities.
Such pairs of
features in HRF-spectra represent the most general form
of amplitude modulation (see, \eg \mbox{paper-I}
and Asgekar 2002); 
however an additional weak phase modulation cannot
be ruled out. There is no clue as yet about the
order of the aliasing of the primary modulation, although,
as we will discuss later, the apparent modulation frequency 
is more likely to be 0.53 than 0.47\,c/$P_1$. 

In addition to the primary fluctuation features, in 
Figure \ref{fig:0834D216_hr} we also find another pair of
features at $0.463\pm 0.069$\,c/$P_1$ flanking the \mbox{feature-1}.
Here, the side-tone product close to 0.4\,c/$P_1$ is
clearly seen (a frequency of $0.390 \pm 0.001$\,c/$P_1$),
whereas the other (upper side-tone product) falls very close to
the `alias' component (at 0.53\,c/$P_1$) of the primary 
fluctuation feature.
These features suggest the presence
of a slower, and most likely an amplitude, modulation on 
the primary fluctuation.
The mean period of this slower modulation is estimated 
to be ~$14.5\pm\,1.0\,P_1$,
where the larger uncertainty reflects uncertainties in the estimation of
frequency of the upper side-tone product located near~0.53\,c/$P_1$. 
So, we observe a primary fluctuation which is modulated by
a slower variation/fluctuation.
Overall, this behavior is qualitatively similar
to pulsar \mbox{B0943+10}.
With this, we also note that there is an excess of power
at $0.073\pm 0.007\,c/P_1$ (a period of $\sim13.7\pm1.4\,P_1$)
in the LRF-spectrum
computed from sequence~B (see Figure~\ref{fig:0834B215_lr}b).

In most parts of sequence~C, we find the same fluctuation spectral feature
as seen in sequence~A. However, in a small section of the C~sequence,
we notice an extreme situation where the alternate-pulse modulation
feature (i.e. feature I) becomes very weak and the dominant modulation
appears to be the slower one. This is clearly evident from the pulse
sequence (Figure \ref{fig:0834A308_proflr}a), where strong emission is 
encountered about every 15 rotation periods. The fluctuation spectrum of
this section of the data (Figure \ref{fig:0834A308_proflr}b) also reveals a
strong feature at ~0.06\,c/$P_1$ that directly relates to the slow and deep 
modulation at a period of $15.0\pm\,0.8\,P_1$.
Not surprisingly, this period is consistent with that suggested by the 
side-tones seen in the B~sequence.

The three data sets presented here display almost a complete
variety in the relative dominance of the two modulation features,
and provide a remarkably consistent picture of their origin.

\noindent{\section{Aliasing, the circulation time, and drift direction}}

We now fold the pulse A \& B~sequences separately at their 
apparent primary fluctuation periods ($P_3$ tentatively identified as 
1.856$P_1$ \& 1.859$P_1$, respectively). The folded time series are
displayed in Figure~\ref{fig:0834B215_mf1}a,b.

The phase of the modulation (not shown) 
varies very slowly with pulse longitude and is not
well defined at the centre of the profile. The intrinsic phase pattern is 
also likely to be smeared by a 5-bin smoothing we performed to improve the
visibility of the otherwise weak pulses.
The apparent average rate of this phase variation is small ($\le 4\,\deg/\deg$),
consistent with the primary feature being associated with
an amplitude modulation. 

The folded sequences also show that the modulation
patterns under the two components of the average profile
are possibly shifted in phase with respect to each other
(Figures~\ref{fig:0834B215_mf1}a,b), but only slightly.
The exact delay is difficult to measure due to the very patchy
nature of the intensity patterns. However, in our estimation
the magnitude of the apparent relative delay between the
modulations is within about 0.1~of
the modulation cycle, or ~0.2\,$P_1$, corresponding to the
longitude separation of the peaks of the two well-defined
components in Figure~\ref{fig:0834B215_mf1}b.

The relative delay in the modulation under the two
components can be expressed, in general, as
\begin{equation}
\delta\zeta = (X + m)\,, with\,\,m=...-2,-1,0,1,2...;
\label{eq:0834_modph}
\end{equation}
where, $X$~is the apparent phase (or delay) difference between
the modulations under the two components expressed as a
fraction of the modulation period, and the integer $m$ accounts
for the possible undetected integral cycles of the modulation
phase as well as asserts the sense/direction of the drift (when combined
with $X$). For the folded version of sequence~A shown, $X$=0.1,
and $X$=-0.1 if the data were to be folded at
$P_3$=2.169\,$P_1$. For the B sequence, we estimate
$X$= -0.05 for the presently assumed $P_3$, and
$X$ = 0.05 for the corresponding longer $P_3$
(Figure~\ref{fig:0834B215_mf1}).

\begin{figure*}
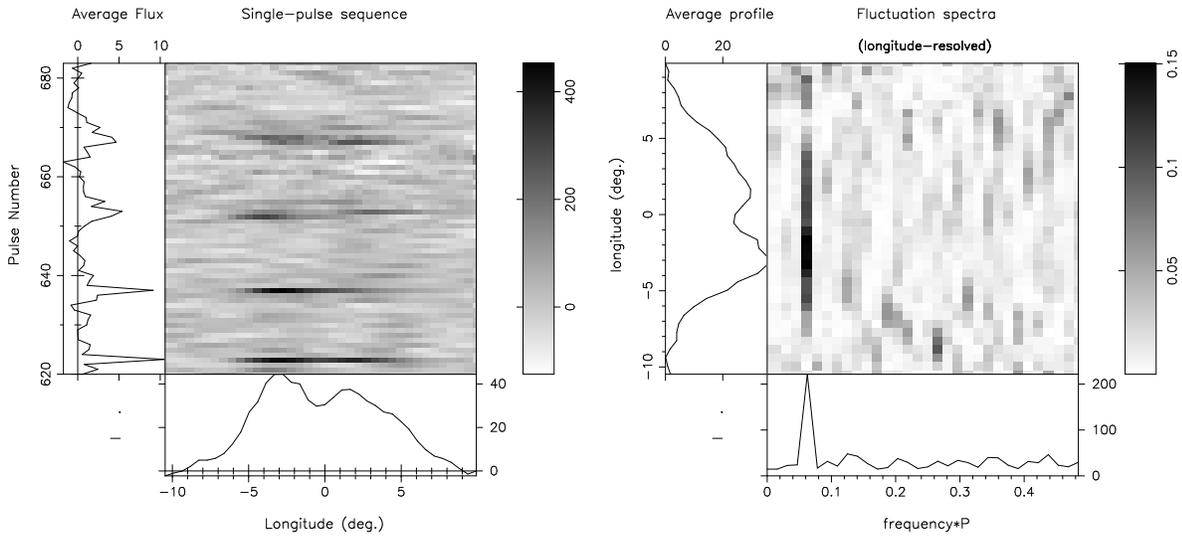

\begin{center}
\begin{tabular}{@{}lr@{}}
{\mbox{\epsfig{file=ME651_FIG_3L.ps,width=7.0cm,angle=-90}}}&
{\mbox{\epsfig{file=ME651_FIG_3R.ps,width=7.0cm,angle=-90}}}\\
\end{tabular}
\end{center}
\caption[]{
(Left) A 64-pulses long section of the C~sequence
shown in gray-scale; and 
(Right) longitude-resolved spectrum of this section
of the sequence. 
Note the strong feature at~0.07\,c/$P_1$ and the near absence
of modulation feature at 0.46\,c/$P_1$.}
\label{fig:0834A308_proflr}
\end{figure*}

\begin{figure*}
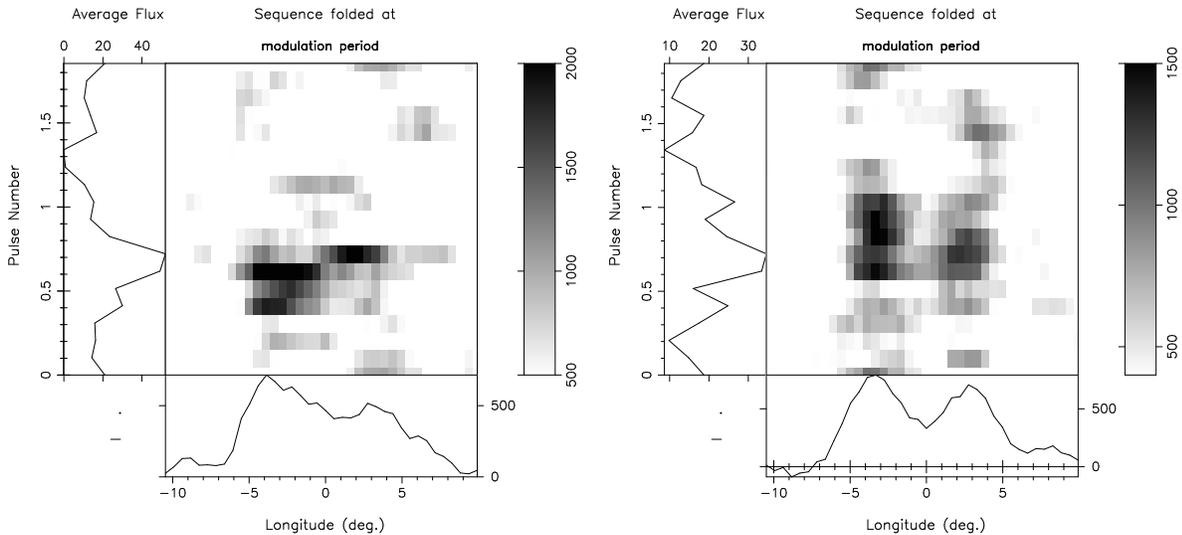

\begin{center}
\begin{tabular}{@{}lr@{}}
{\mbox{\epsfig{file=ME651_FIG_4L.ps,width=7.0cm,angle=-90}}}&
{\mbox{\epsfig{file=ME651_FIG_4R.ps,width=7.0cm,angle=-90}}}\\
\end{tabular}
\end{center}
\caption[]{(Left) Sequence~A folded at a modulation
period of $P_3$ = 1.856\,$P_1$.
The relative delay in the
modulation peaks under the two components is evident. A similar
pattern would be seen even if the other possible $P_3$
(\ie,~2.169$\,P_1$) were to be used for folding; then, any
``drift" (or relative delay) would appear in the opposite sense.
(Right) Sequence~B folded at a $P_3$ of
1.859\,$P_1$.
}
\label{fig:0834B215_mf1}
\end{figure*}

We interpret the relative delay between modulation under the two
components as caused by the `passage' of the same
emission entity through our sightline as the polar-emission
pattern rotates steadily around the magnetic axis.
We can write the geometrical relationship between the rotational
longitude $\delta\varphi$ and magnetic azimuth angle
$\delta\theta$ as,
\begin{equation}
\delta\theta\,= \,sin^{-1}\left[sin(\alpha+\beta)
                sin(\delta\varphi)/\mid sin(r_{\varphi})\mid\right],
\label{eq:mod_diff_fine}
\end{equation}
where $r_{\varphi}$, the associated magnetic latitude at the component
longitudes, can be computed using the following well known relation
\begin{equation}
sin^2(r_{\varphi}/2) = \,sin^2(\delta\varphi/2) sin(\alpha+\beta) sin(\alpha)
                 + sin^2(\beta/2)
\label{eq:rho_def}
\end{equation}
A rotational longitude separation of the component peaks of 
~$\Delta\varphi=6.2\deg$ (i.e. $\Delta\varphi=\pm3.1\deg$)
implies a magnetic longitude separation
$\Delta\theta\simeq 50\deg$ (i.e. $\delta\theta\simeq \pm25\deg$)
around the magnetic
axis of the star\footnote{Here, we have used $\alpha=30\deg$
and $\beta=-3\deg$ in the above estimation. The 
value of $\alpha$ as well as the sign of $\beta$ are
uncertain. However, the magnitudes of $\alpha$ and $\beta$
are consistent with the PA-sweep rate reported by 
\mbox{Stinebring \etal (1984)}. An independent detailed assessment
based on profile evolution across a wide frequency range and PA
sweep rate (14.5 $\deg/\deg$) gives a somewhat different viewing
geometry, with $\alpha$ = 45$\deg$ and $\beta$ = -2.8$\deg$ (Rankin, 
private communication). The presently adopted geometry implies a
shallower PA sweep (~10 $\deg/\deg$) that was arrived at based on
the `closure' technique results discussed in the next section.}. 
We can now directly estimate the circulation time of the pattern using
eq.~(\ref{eq:mod_diff_fine}). Assuming that the polar cap
periphery is roughly circular (see, for example, \mbox{paper-I};
Mitra \& Deshpande 1999), the circulation time is given by
\begin{equation}
\hat{P}_3 = \left(\frac{360\deg}{\Delta\theta}\right)
		{\delta\zeta} P_3
\label{eq:p3_one}
\end{equation}
or, with eq. (1),
\begin{equation}
\hat{P}_3 = \left(\frac{360\deg}{\Delta\theta}\right)
	      (X\,+ m)\,P_3\;\,\, (m=\ldots,-1,0,1,2,\ldots),
\label{eq:p3_two}
\end{equation}
where $\Delta \theta$ is the magnetic-longitude separation
(in degrees) of the two component peaks.

\begin{figure}
\centerline{\epsfig{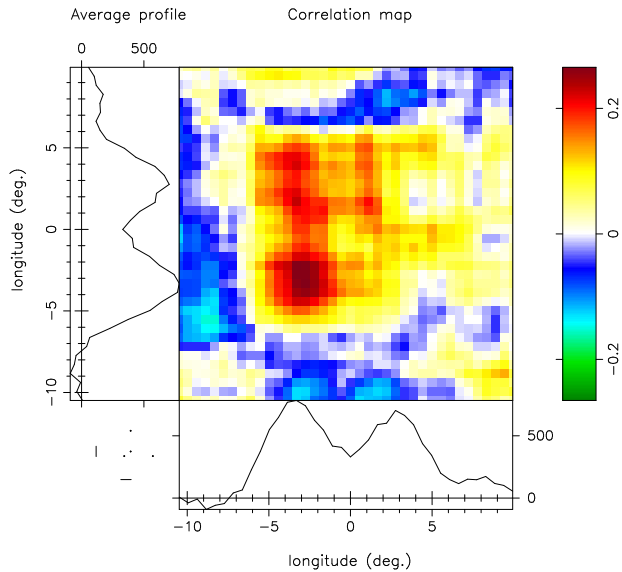}}
\caption[Longitude-longitude correlation map of \mbox{B0834+06}: $\mu = 2$]
{Longitude-longitude correlation of intensity fluctuations
at a relative delay of 2 pulse periods, computed for B~sequence data 
(of some~350 pulses). The intensity fluctuations at each longitude bin, 
shown along the horizontal
axis, are cross-correlated separately with the fluctuations seen two pulse period later at 
every longitude bin indicated along the vertical axis.
The central panel gives this correlation map according to the
color scale at the right.
The bottom and the left panels show the average profiles for the original B~sequence
and for the delayed version of it, respectively.
The colour-scale used to display the correlation coefficients
in the map is shown on the right.
The observed correlation map is
asymmetric \wrt\,\,to the diagonal. The sense of the
asymmetry indicates a
subpulse ``motion" from earlier to later longitudes.}
\label{fig:0834_llc}
\end{figure}

If the observed amplitude modulation is generated by  
the steady rotation of a stable pattern around the magnetic
axis, it should produce fluctuations having
a period equal to the circulation time.
The estimates for the two sequences and different
values of~$m$ are to be compared with our estimates of the circulation time
as suggested by the slower modulation period (i.e., ~14.8\,$P_1$ or so).
It can be shown that the compatible values of $m$ are +1 and -1.
Given the uncertainties in the various relevant parameters, it is not possible 
as yet to conclusively estimate the sign of $m$.

To help resolve this issue, we have chosen intensity sequences
corresponding to two longitudes intervals symmetrically placed around the
central (zero) longitude and with a separation of 6$\deg$ in order to
examine their cross-correlation. In every one of the three sequences
(i.e., A, B, \& C),
we find that the cross-correlation has a peak when the earlier longitude
data is delayed by two pulse periods w.r.t. that at the later  longitude,
in clear contrast with the other way around.
This clearly implies that $m$=+1, and that the drift direction is from
earlier longitude to later--- {\it i.e.}, opposite to the 
pulsar rotation direction.

A further related confirmation of the drift direction
(relative to the pulsar rotation) comes from 
longitude-longitude correlation maps of
pulse sequences over delays of integer number of pulse
periods (such an analysis was first used by 
Popov \& Sieber 1990). The correlation
map computed for sequence~A is displayed in
Figure~\ref{fig:0834_llc}, where 
the estimated (normalized) correlation between fluctuations in the intensity 
at a range of longitudes (marked along the horizontal axis) and
the fluctuations  observed two pulse-period later (at the various longitudes
marked along the vertical axis) is shown.

Significant auto-correlation (i.e., along the +45$\deg$--slope line)
is apparent at this delay of
$+2\,P_1$, which is related to the primary modulation
periodicity seen in this pulsar. The cross-correlation
map, however, is not diagonally symmetric (about the +45$\deg$--slope line)
for a delay of two rotation-periods,
that is, the delayed versions of later-longitude
sequences do not correlate with those at earlier longitudes.
This we interpret as subpulse ``motion" 
from earlier to later rotational longitudes,
\mbox{\ie}, positive drift.

Thus, we have identified an internally consistent configuration of the underlying
circulating subbeam system  
responsible for the observed modulation. Based on various different
signatures, we have indeed been able to estimate the circulation time, and find 
that $14.8\pm 0.8\,P_1$ is a value consistent with our observed sequences.
It is worth pointing out that the actual circulation time is
likely to be somewhat shorter than the above value (by typically 1 $P_1$),
given the typical frequency of nulls.
The drift direction is also determined above, and the sense of the PA sweep
is known from the existing polarization data.

The alias ambiguity of the primary modulation feature, however, has yet to be resolved;
but this is really only of academic interest, since our further
investigations based upon emission mapping, \etc, do not depend
on its resolution. As we will mention later, it
is possible to simply look at the average sequence over several 
circulation times and estimate the apparent spacing between
successive sub-beams. 

None of our derived estimates so far depend crucially on 
an accurate knowledge of the viewing geometry, although it was useful in
determining the value of~$m$, and the geometry will be an explicit
ingredient in the cartographic transform to be used for
emission mapping. We note that the available polarization measurements
on this pulsar do not effectively constrain the viewing geometry, especially
the sign of $\beta$.

\noindent{\section{Viewing Geometry of B0834+06}}
\label{sect:0834_geom}

B0834+06, based on its {\it double} profile, is a member of
the ``D"-class according to the Rankin classification.
The emission of this pulsar is believed to arise
from the inner conal ring, and the geometrical parameters were
derived accordingly (Rankin 1993). 

The `inverse' of the cartographic transform provides
a powerful ``closure" path to verify and refine the input
parameters used to produce the polar map (see \mbox{paper-I}
and Deshpande~2000, for a detailed discussion of this point).
We wished to reconfirm the drift direction and ``tune" our
estimate of the circulation time. We also wanted to further
constrain the assumed viewing geometry using this approach, if
possible. To achieve this we carried out a grid search, where
we varied the values of $\alpha$, $\beta$ and $\hat{P_3}$ over a
reasonably wide range and alternated the drift direction 
as a parameter. For each different combination of the
parameters in the search, we constructed an average-polar
map from the observed sequence. The inverse
transform was used to generate an artificial
sequence from this trial polar emission pattern. We then
cross-correlated (longitude by longitude) 
the artificial sequence with the natural
(observed) one. The parameter combination for
which the procedure yielded a higher cross-correlation
coefficient, averaged over the longitudes, was
considered a better indicator of the true values.

For B0834+06, the closure technique involving the inverse
transform turns out to be rather insensitive to changes
in the values of
$\alpha$ and $\beta$ ($\alpha$ being larger for 
\mbox{B0834+06} than for \mbox{B0943+10}) and
the sign of $\beta$ was not constrained (see, Asgekar 2002).
The technique was, however, sensitive to the circulation
time and the drift direction, which allowed us to
constrain these parameters. This, in a way, also
meant that our resulting polar emission maps were
rather insensitive to uncertainties in
$\alpha$ and~$\beta$. 
In what follows, we use the following set of refined
parameters:
$\hat{P_3} =  14.84 P_1$;
$\alpha =  30\deg$;
$\beta =  -3\deg.0$,
with the drift direction same as that of the
pulsar rotation. The resolution of
the drift direction is then tied to the ambiguity
in the sign of $\beta$, and a comparison with Ferraro's
theorem (Ruderman 1976) is possible after an
unambiguous determination of~$\beta$.

\noindent{\section{Polar Maps and Movies}}
\label{sect:0834_maps}

We created the polar emission maps for the three
pulse sequences discussed earlier using the cartographic
transform mapping technique of Deshpande \& Rankin
(DR99, paper-I).
Also, series of maps, each from successive short
subsequences (of duration~$\ge\,\hat{P}_3$) 
and with a large fractional overlap with
the neighboring ones, were generated. The resulting
images were viewed in a slow `movie'-like fashion.
Since a polar map over one circulation time for B0834+06
doesn't sample the entire polar cap adequately, each movie frame
was made using \mbox{30 pulses}
(roughly twice the circulation time $\hat{P_3}$).
We further smoothed
this pattern over $10\times 10$~points in order to
reduce the noise contribution, the smaller number of sub-beams 
making such smoothing practical.
The movies allowed us to
study variation of the polar emission pattern
over the duration of the observation.
Here, we present our maps and the insights gained
by studying the movies. 

\noindent{\subsection{Sequence-A ($5^{th}$February 1999)}}

\indent The polar map for this sequence is shown in
Figure~\ref{fig:0834B215_polar}a. Only a couple of
distinct sub-beams are seen in the average polar map,
whereas short averages corresponding to the frames
of the movie show us more distinct sub-beams that
fluctuate around their mean positions and in brightness.
Patterns of strong,
distinct sub-beams appear on the scales of $\sim2\hat{P}_3$,
giving rise to weak (Q-value $\la 100$) primary
fluctuation features in the spectra. On the whole a
few sub-beams dominate the maps over
short intervals, similar to the case of
\mbox{sequence~B} below. 
Such `prominences' last only about two circulation
times, where after those
sub-beams become weak and some others more prominent.

The bright sub-beams that dominate the emission in a
movie frame appear to be broader than the average.
Assessing the significance of this impression is difficult given
both the smoothing and the noise
present, but at times two adjacent bright
sub-beams appear to  touch each other.

The bright, isolated emission patches near the outer
periphery of the region mapped are due to occasional
extraordinarily bright single pulses in our sequence.
We discuss these below.

\noindent{\subsection{Sequence-B ($6^{th}$ February 1999)}}

	This sequence shows the sharpest primary
fluctuation feature which has 
a $Q$-value~$\sim100$. Its polar map,  
presented in Figure~\ref{fig:0834B215_polar}b,
shows a system of distinct sub-beams.
Similar to the observations of \mbox{B0943+10},
the brightness of individual sub-beams appear to
fluctuate with time. In fact, they vary
by factors of almost 5 between certain short averages.
At any given time a few sub-beams dominate
the emission and overshadow emission from the rest
of the polar cap. The individual subbeams here are generally
less stable compared to those of
\mbox{B0943+10}, and show an increased
tendency to fluctuate around their mean positions 
---which may explain
the lower Q value of this star's primary fluctuation feature.

We would like to attract the reader's attention to the part
of the map where five roughly equi-spaced sub-beams span
almost exactly half of the magnetic longitude range. It
is clear from this configuration that the true $P_3$ is about
1/8$^{th}$ of the circulation time, {\it i.e.} less than 2$\,P_1$.
This, then, finally 
appears to resolve the aliasing issue.
The LRF feature at~0.47 $c/P_1$ is indeed an alias of the true
modulation frequency of $\sim 0.53$ $c/P_1$, making the unaliased
$P_3$=1.859 $P_1$. This, in turn, implies the typical number of subbeams
to be 8, a number also consistent with the estimates of magnetic
azimuthal separation of the subbeams (as in section 4). Nonetheless,
the subbeams spacing is not as uniform as in \mbox{B0943+10}.

\noindent{\subsection{Sequence-C ($5^{th}$February 2002)}}

The short 64-pulse sequence mapped here exhibits a 
deep periodic modulation that reflects the 
circulation cycle directly. Only a single dominating emission region is
apparent.
The streaky and highly elongated
shape of the active region is the result of poor sampling
and inadequate averaging across the map area. This
extreme situation is rare as well as rewarding, in the sense that it
permits us a direct estimation of the circulation period.\\

\begin{figure*}
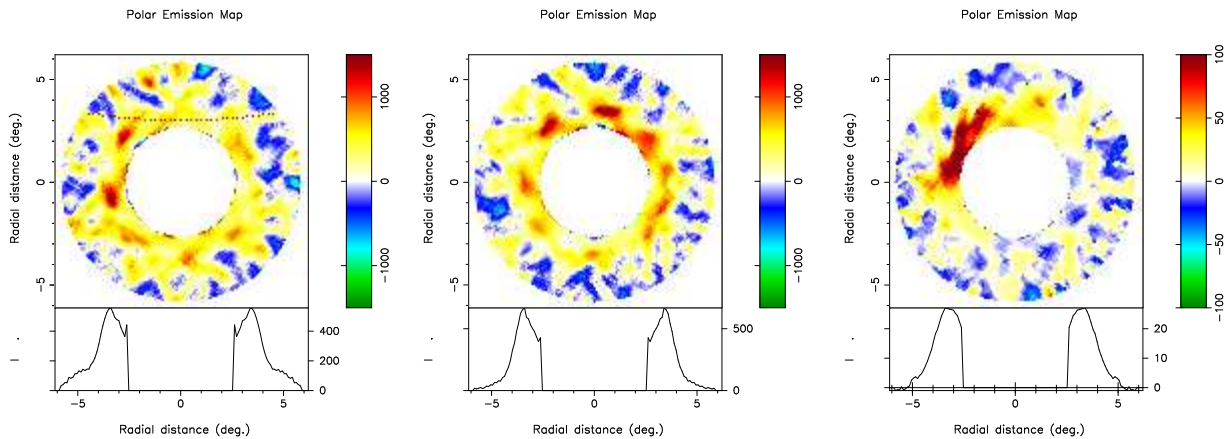

\begin{center}
\begin{tabular}{@{}lr@{}rr@{}}
{\mbox{\epsfig{file=ME651_FIG_6L.ps,width=5.7cm,angle=-90}}}&
{\mbox{\epsfig{file=ME651_FIG_6C.ps,width=5.7cm,angle=-90}}}&\hspace*{12pt}
{\mbox{\epsfig{file=ME651_FIG_6R.ps,width=5.7cm,angle=-90}}}\\
\end{tabular}
\end{center}
\caption[Average polar maps of emission for \mbox{B0834+06}
\mbox{sets-A, B \& C}]
{
(Left) Average polar emission map for \mbox{B0834+06}
\mbox{sequence-A}
made with 250~pulses. The sub-beams are less distinct and 
not equispaced, which contributes to the low Q-value of
primary fluctuation. Considerable averaging has been carried out 
to improve the map's signal-to-noise ratio. The string of
dots at the top shows the track sampled by our sight-line.
(Center) Average polar emission map for 
\mbox{sequence B} using $\sim350$ pulses. The 
sub-beams intensities here show a systematic variation with azimuth, a
property that manifested itself as the side-tones in
the fluctuation spectra of this sequence.
(Right) A map made using 64 pulses from sequence C.
The streaky appearance results from inadequate sampling and/or
averaging of the mapped region. Note the single prominent
emission region and almost complete lack of
emission over the rest of the polar region.
The colour-scales used to display the intensities
in the polar maps are shown on the right-side of respective
maps. The apparent --ve intensities are a consequence of 
noise dominance in the off-pulse region.}
\label{fig:0834B215_polar}
\end{figure*}

\noindent{\section{Discussion}}

In the preceding sections we have presented our analysis
of the fluctuation properties apparent in the decametric
emission of pulsar \mbox{B0834+06}, as well as our
efforts to understand the underlying emission pattern
responsible for them.

\mbox{B0834+06} is the {\it only} other pulsar
(apart from \mbox{B0943+10}) for which we could
estimate the circulation time of the polar emission
pattern reliably using its fluctuation properties
and viewing geometry. This estimation is based
solely on our decameter observations. We have made
polar emission maps from our \mbox{34.5 MHz} sequences
using the estimated $\hat{P}_3$ and drift direction,
and those were presented in the earlier section.
In carrying out our analysis, we followed a general
strategy for the estimation of the circulation time
of the polar emission pattern. This approach could be
applied to other pulsars as well. We initially computed 
LRF- and HRF-spectra for this pulsar, and have shown
that
\begin{itemize}
\item The primary modulation has a time
delay between the two components of the star's double 
profile.
\item We argued that the amplitude
fluctuations of the two components result from the
`passage' of the same emission entities as the 
polar emission pattern revolves progressively around
pulsar's magnetic axis. This, for the component
separation of $50\deg$ in magnetic longitude and
the apparent modulation delay, suggested that the circulation
time ($\hat{P}_3$) is ~$\sim14.9$\,$P_1$.
\item The HRF-spectrum, computed using
\mbox{set-B}, exhibits sidebands around the primary
modulation (see Fig.~\ref{fig:0834D216_hr}).
These sidebands imply a slower periodicity modulating
the primary modulation feature, having a period
of $\sim14.5\pm0.8$\,$P_1$. We notice discernible power
in the LRF-spectrum of the same sequence
(Fig.~\ref{fig:0834B215_lr}) at a frequency close
to 0.073\,c/$P_1$ ({\it i.e.} corresponding to a period of~$\sim14 P_1$).
\item The rewarding 64-pulse sequence (set-C)
allowed a direct estimation of the circulation period,
based on the low frequency feature in its fluctuation
spectrum.  This period estimate ($\sim15 P_1$) is
consistent with the values determined from the other
two sequences.

\item We believe that these observations conclusively
demonstrate the circulation time to be $\sim14.9$\,$P_1$.
\item The estimates of the viewing geometry for 
this pulsar are uncertain. We carried out a
search to refine the viewing geometry and other
parameters of transform using the inverse of the
cartographic transform. This search constrained
the circulation time (now refined
to be $\hat{P}_3$\,=\,14.84\,$P_1$) and the drift
direction of the pattern relative to the pulsar
rotation, both of which are crucial for our mapping technique.
However, $\alpha$ and $\beta$ were relatively poorly constrained
in the case of this pulsar (B0834+06) and with this approach.
\item A system of discrete sub-beams in steady rotation
around the magnetic axis was found responsible for the
observed pulse-to-pulse fluctuations. Because
of the rather central traverse of the observer's
sightline, the circulation of the sub-beam system manifests as
amplitude modulations, not as a subpulse drift.\\
\end{itemize}
The emission sub-beams of \mbox{B0834+06}
are discrete. They are not uniformly spaced in 
magnetic azimuth, and are irregular in form.
The sub-beams appear to be wide in the
radial direction, \mbox{with
$\Delta\rho/\rho\ga25\%$}.
We had concluded
from our analysis of \mbox{B0943+10} that the regions
of radio emission correspond to a well-organized
system of plasma columns. In \mbox{B0834+06}
we observe less order 
and more fluctuation in 
subbeam locations and brightness. This increased
jitter in subbeam position, in addition to
their non-uniform spacings, could result in the
lower Q-values observed in the star's fluctuation features.
It was not possible to account for the null pulses,
which could further affect the modulation~$Q$.

We also made maps from successive subsections
of the sequences and viewed them in a `movie'-like
fashion. The variation, as noted above, in their relative locations
and brightness is even more apparent within and across these maps.
Although it is difficult to quantify these due to limited sensitivity,
it is clear that usually only a couple of bright sub-beams 
seem to dominate the polar emission of this pulsar. 

The isolated, bright emission patches close to the
periphery appear due to a handful of strong pulses
in our data and these areas receive negligible emission
from a majority of the
pulses. The origin of such intense pulses is
unclear, however it may correspond to   
a {\it Q-mode} activity similar to that observed in pulsar \mbox{B0943+10}
(see \mbox{paper-I}).

The significance of our analysis also stems from the fact
that pulsar \mbox{B0834+06} is {\it different} from
\mbox{B0943+10} in its physical characteristics, its
viewing geometry and hence in the nature of the apparent
modulation it exhibits.
The single-pulse fluctuations here mainly 
appear as modulations of its component amplitudes, unlike the
drift patterns (or phase modulation) observed in \mbox{B0943+10}. 
Nonetheless we were able to relate such a general
form of modulation to a well-defined
rotating pattern in the polar emission region. This
supports the picture that a set of discrete emitting sub-beams
circulating within a hollow
cone of emission, as envisioned in R\&{S}, produces the
single-pulse fluctuations in pulsars. The specific 
characteristics of the apparent fluctuations, e.g. whether 
it is an amplitude and/or phase modulation,
would then depend mainly on the observer's viewing geometry. 

We now wish to quantitatively compare our current results
as well as those obtained previously on \mbox{B0943+10}
with theoretical models.
We tabulate all the relevant parameters of these two
pulsars in Table~\ref{tab:compare}, where the magnetic
field estimates are derived from their timing behavior.

R\&{S} express the circulation time of the discrete
spark pattern as,
\begin{equation}
\hat{P}_3/P_1\,= K_0 (B_{12}/P_1^2),
\label{eq:P3_cap}
\end{equation}
where $B_{12}$ is the average magnetic field (in units of 10$^12$ gauss)
in the gap and
$P_1$~is the pulsar rotation period in seconds. The constant
$K_0$ (= 5.6 according to R\&{S}) is considered
to be a very shallow function of conditions local to
the emission region.
\begin{table*}
\begin{center}
\begin{tabular}{c|c|c|c|c|c|c|c|c|c|c}
\hline
Pulsar&$\alpha$&$\beta$&$B_{12}$&$P_1$&number of&$P_3$&$\hat{P}_3$ (R\&{S})&$\hat{P}_3$(G\& S)&$\hat{P}_3$ (obs)&$\hat{P}_3$\\
&&&&${\it s}$&sub-beams&(in ${\mathit P_1}$)&(in ${\mathit P_1}$)&(in ${\mathit P_1}$)&(in ${\mathit P_1}$)&{\it (s)}\\
\hline
B0943+10&$11\deg.64$&$-4\deg.31$&2&1.097&20&1.848&9.3&36.95&36.95&40.57\\
\hline
B0834+06&$30\deg$&$-3\deg$&3&1.273&8&1.855&13.1&103&14.84&18.9\\
\hline
\end{tabular}
\caption[Comparison between the observed circulation time and theoretical
estimates of R\&{S} model]{The various pulsar parameters for
pulsars \mbox{B0943+10} and \mbox{B0834+06} are listed. We also
compare between the observed circulation time and the theoretical
estimates (see text).}
\label{tab:compare}
\end{center}
\end{table*}
From eq.~\ref{eq:P3_cap} we write,

\begin{equation}
\left( \frac{ ({\hat{P}_3})_{0943+10} }{ ({\hat{P}_3})_{0834+06} } \right)_{estimated}
= K
\left( \frac{ ({\hat{P}_3})_{0943+10} }{ ({\hat{P}_3})_{0834+06} } \right)_{observed}
\end{equation}
such that the parameter $K$ accommodates a possibility 
that $K_0$ in eq.~\ref{eq:P3_cap} may not be
a {\it constant}.
K=1 would imply that the proportionality constant $K_0$ (irrespective of its 
precise value) in the
eq.~\ref{eq:P3_cap} remains unchanged from one pulsar
to another. We, however, find the value of K to be~0.29.
This suggests that $K_0$ is likely to have
significant dependence on the pulsar parameters, particularly
the estimated surface magnetic field and the inclination
angle~$\alpha$ (see Table~\ref{tab:compare}). 
In order to meaningfully assess the exact nature of these and other
possible dependences 
we would need $\hat{P}_3$ estimates  to become available for
several other pulsars.

Gil \& Sendyk (2000; hereafter G\&S) have advanced a
modified-R\&{S} spark model for radio pulsars. They have
examined the estimates of subpulse drift and the details
of polar emission maps from DR99 and \mbox{paper-I}. They
claim to exactly reproduce such a behavior from their model,
and the observations of B0943+10 allowed them to fix certain
model parameters. We computed their predicted circulation
time for \mbox{B0834+06} given the parameters in
Table~\ref{tab:compare}.
According to G\&S model \mbox{B0834+06}'s circulation time should be
$\sim103\,P_1$ (for a ``filling factor" of 0.1; and $\sim78\,P_1$ for
a filling factor computed according to their eq. 4), almost an order of 
magnitude higher than we have estimated from our observations. 
On the other hand, interestingly,
the circulation time predicted \mbox{B0834+06} based on the R\&S model
appears to be consistent with that observed. However, according to
Gil, Melikidze \& Geppert (2003; GMG2003) the original R\&S derivation is in
error (by a factor of 4, in $K_0$). 
The estimates for the circulation times in the two cases,
based on their (i.e. GMG2003's) recent detailed formulation 
of parameters associated with the ExB drift, 
show good agreement with those observed. 

Our analyses were carried out using observations obtained at \mbox{34.5 MHz}
alone, and we do not have any polar emission maps made using
single-pulse sequences at meter-wavelengths for comparison.
We believe that such a structure of discrete sub-beams is valid
even at higher frequencies. It would be possible to better account
for the ``null" pulses and their effect on the subbeam motion
using the higher sensitivity of meter-wavelength data. Gil (1987)
has reported the presence of strong orthogonal polarization
modes (OPMs) in this pulsar. Hence, this pulsar may
provide a useful context for studying possible causal relationships 
between the ``null"
pulses, OPMs and amplitude modulations.

\noindent{\section{Conclusions}}

Our observations of \mbox{\mbox{B0834+06}} at 35~MHz 
show that the regular amplitude modulations observed
in this pulsar are due to a system of discrete sub-beams 
rotating more or less steadily around the magnetic axis.
This implies that the general amplitude modulations
and regular ``drifting subpulses" observed in some
pulsars (e.g. Backer 1973) occur due to the same underlying reason. The nature
of the modulation depends on the geometry; a shallow cut by
the locus of the observer's sightline leading to a conal single
profile and subpulse drift, whereas a more central cut would
lead to general amplitude modulations of the components of the
average profile. The subbeam positions and intensities are
not as stable as those observed in \mbox{B0943+10} 
and we observed less order in our maps of \mbox{B0834+06}.
We compared our estimates of the circulation time for
pulsars \mbox{B0834+06} and \mbox{B0943+10} with the
predictions of two specific theoretical models. We find that
the models fail to predict the circulation 
time in a consistent manner for the two pulsars.
Future similar data on other pulsars should help clarify possible
dependences of circulation period on magnetic inclination angle and
other parameters.\\

\section*{Acknowledgments}
We thank the staff at the Gauribidanur Radio Observatory,
in particular \mbox{H. A. Ashwathappa}, \mbox{C. Nanje Gowda},
and \mbox{G. N. Rajasekhar}, for their invaluable help during
observations. We are thankful to V.~Radhakrishnan
and Joanna Rankin for many fruitful discussions
and their critical comments on the manuscript. We also thank our referee, 
Janusz Gil, for his useful comments. This research has made
use of NASA's ADS Abstract Service.

\beb
\bi Asgekar, A. 2002, Ph. D. thesis, Indian Institute of Science, Bangalore.
\bi Asgekar, A. \& Deshpande, A. A. 1999, Bull. Astr. Soc. India, 302, 27
\bi Asgekar, A. \& Deshpande, A. A. 2000, ASP Conference Series, Vol. 202;
Proc. of IAU Colloquium \# 177, (San Francisco: ASP), Ed. M. Kramer,
N. Wex, and R. Wielebinski, pp. 161
\bi Asgekar, A. \& Deshpande, A. A. 2001, MNRAS, 326, 1249 (\mbox{paper-II})
\bi Backer, D. C. 1973, ApJ, 182, 245
\bi Deshpande, A. A. 2000, ASP Conference Series, Vol. 202;
Proc. of IAU Colloquium \# 177, (San Francisco: ASP), Ed. M. Kramer,
N. Wex, and R. Wielebinski, pp. 149
\bi Deshpande, A. A. \& Radhakrishnan, V. 1994, JApA, 15, 329
\bi Deshpande, A. A., Ramkumar, P. S., Chandrasekaran, S., Vinutha, C. \& Prabu, T. 2004, in preparation. 
\bi Deshpande, A. A. \& Rankin, J.M. 1999, ApJ, 524, 1008 (DR99) 
\bi Deshpande, A. A. \& Rankin, J.M. 2001, MNRAS, 322, 438 (\mbox{paper-I}) 
\bi Deshpande, A. A., Shevgaonkar, R. K. \& Sastry, Ch. V. 1989, JIETE, 35(6), 342
\bi Drake, F. D., Craft, H. D., Jr. 1968, Nature, 220, 231
\bi Gil, J. A. 1987, ApJ, 314, 629
\bi Gil, J. A. \& Sendyk, M. 2000, ApJ, 541, 351
\bi Gil, J. A., Melikidze, G. I. \& Geppert, U. 2003, A\&A, 407, 315
\bi Gupta, Y., Gil, J. A., Kijak, J. \& Sendyk, M. 2004, A\&A, 426, 229
\bi Lyne, A. G. \& Ashworth, M. 1983, MNRAS, 204, 519
\bi Mitra, D. \& Deshpande, A. A. 1999, A\&A, 346, 906
\bi Popov, M. V. \& Sieber, W. 1990, Sov Astron, 34, 382
\bi Rankin, J. M. 1986, ApJ, 301, 901
\bi Rankin, J. M. 1993, ApJ, 405, 285
\bi Rankin, J. M., Suleymanova, S. A. \& Deshpande A. A. 2003, MNRAS, 340, 1076
\bi Ritchings, R. T. 1976, MNRAS, 176, 249
\bi Ruderman, M. A. \& Sutherland, P. G. 1975, ApJ, 196, 51 (R\&{S})
\bi Ruderman, M. A. 1976, ApJ, 203, 206
\bi Slee, O. B. \& Mulhall, P. S. 1970, \PASA, 1, 322
\bi Stinebring, D. R., Cordes, J. M., Rankin, J. M., Weisberg, J. M. \& Boriakoff, V. 1984, ApJS, 55, 247
\bi Sutton, J. M., Staelin, D. H., Price, R. M. \& Weimer, R. 1970, ApJ, 159, L89
\bi Taylor, J. H., Jura, M. \& Huguenin, G. R. 1969, Nature, 223, 797
\bi van Leeuwen, A. G. J., Stappers, B. W., Ramachandran, R. \& Rankin, J. M. 2003, A\&A, 399, 223
\eeb
\end{document}